\documentclass[aip,amsmath,amssymb,reprint]{revtex4-2}
\usepackage{graphicx}
\usepackage{dcolumn}
\usepackage{bm}
\usepackage[utf8]{inputenc}
\usepackage[T1]{fontenc}
\usepackage{mathptmx}
\usepackage{etoolbox}
\usepackage{pifont}
\usepackage{hyperref}
\usepackage[mathlines]{lineno}

\begin{document}

\title{Field-Assisted Sub-Terahertz Spin Pumping and Auto-Oscillation in NiO}
\author{Mingda Guo}
\affiliation{Department of Physics and Astronomy, University of California, Riverside, California 92521, USA}
\author{Ran Cheng}
\email[]{ran.cheng@ucr.edu}
\affiliation{Department of Electrical and Computer Engineering, University of California, Riverside, California 92521, USA}
\affiliation{Department of Physics and Astronomy, University of California, Riverside, California 92521, USA}

\begin{abstract}
Spin pumping converting sub-terahertz electromagnetic waves to DC spin currents has recently been demonstrated in antiferromagnets (AFMs) with easy-axis magnetic anisotropy. However, easy-plane AFMs such as NiO, which are easier to prepare experimentally, are considered to be bad candidates for spin pumping because the N\'{e}el vector oscillation is linearly polarized, placing a major restriction on the material choice for practical applications. Through a case study of NiO, we show that an applied magnetic field below the spin-flop transition can substantially modify the polarization of the resonance eigenmodes, which enables coherent sub-terahertz spin pumping as strong as that in easy-axis AFMs. In addition, we find that an applied magnetic field can significantly reduce the threshold of N\'{e}el vector auto-oscillation triggered by spin-transfer torques. These prominent field-assisted effects can greatly facilitate spintronic device engineering in the sub-terahertz frequency regime.
\end{abstract}

\maketitle

As an emerging frontier in physics and electrical engineering, antiferromagnetic (AFM) spintronics initiated a revived search for electronic devices operating in the terahertz (THz) frequency range, thanks to their unique physical characteristics such as ultrafast dynamics and vanishing stray fields~\cite{baltz2018antiferromagnetic,jungwirth2018multiple,gomonay2014spintronics}. Recently, sub-THz spin pumping~\cite{cheng2014spin,johansen2017spin} has been confirmed experimentally in easy-axis AFM MnF$_2$~\cite{vaidya2020subterahertz} and Cr$_2$O$_3$~\cite{li2020spin}. However, easy-plane AFM materials such as NiO and MnO, which are more abundant in nature and easier to prepare experimentally, are considered to be bad candidates for realizing spin pumping. This is because the dynamics of the N\'{e}el vector is linearly polarized, blamed on the broken rotational symmetry due to the easy-plane anisotropy~\cite{sievers1963far}. Recently, it has been demonstrated that in light of the Dzyaloshinskii–Moriya interaction, spin pumping can be achieved in easy-plane AFM $\alpha$-Fe$_2$O$_3$~\cite{wang2021spin,boventer2021room}. However, what pumps the DC spin current in $\alpha$-Fe$_2$O$_3$ is the weak magnetic moment rather than the N\'{e}el vector~\cite{williamson1964AFM}, so the system effectively functions as a ferromagnet with an operating frequency limited to tens of gigahertz. It is still an open question as to how easy-plane AFM materials can be exploited as a general platform to realize spin pumping in the sub-THz regime, especially at room temperature. Furthermore, as a reciprocal effect of spin pumping, current-induced auto-oscillation in easy-plane AFM materials requires an impractically high current threshold~\cite{cheng2016terahertz,khymyn2017AFM}, because the current-induced torques cannot exhibit any anti-damping effect before they fully counteract the impact of easy-plane anisotropy~\cite{cheng2016terahertz}. Therefore, within our current understanding, simple room-temperature AFM materials such as NiO can hardly be practical choices for ultrafast device applications.

One way out of this dilemma, basing on pure symmetry considerations, is to compensate the broken rotational symmetry caused by the easy-plane anisotropy. Applying a magnetic field along the in-plane easy axis can achieve this goal because the Zeeman interaction bears rotational symmetry around the magnetic field. To confirm whether an applied magnetic field will enable spin pumping in easy-plane AFM materials, it is essential to figure out how the applied field can affect the polarization of the resonance modes [for field strengths below the spin-flop (SF) threshold]. However, this piece of information remains elusive even theoretically, although the dependence of resonance frequencies on an applied field has been well established~\cite{machado2017spin,rezende2019introduction}.

In this Letter, we conduct a case study of NiO---a prototype of easy-plane AFM systems, to demonstrate that a bias magnetic field applied along the in-plane easy axis direction (without inducing the SF transition) can: 1) enable detectable sub-THz spin pumping by substantially modifying the polarization of sublattice spin dynamics; 2) significantly reduce the current threshold triggering auto-oscillation of the N\'{e}el vector by counteracting the influence of easy-plane anisotropy. Our findings pave the way to utilize simple room-temperature AFM materials to perform ultrafast functionalities.

Let us first consider a two-sublattice macrospin model where the two antiferromagnetically coupled magnetic moments are described by $\bm{M}_1$ and $\bm{M}_2$. We choose the coordinate system such that the easy-axis is along $\hat{x}$ and the hard-axis is along $\hat{z}$ (so the $x-y$ plane is the easy plane). In NiO, $\hat{z}$ is the $(111)$ direction and $\hat{x}$ is the $(11\bar{2})$ direction. In terms of the unit vectors $\bm{m}_1=\bm{M}_1/M_s$ and $\bm{m}_2=\bm{M}_2/M_s$ with $M_s$ the saturation magnetization, the free energy density of NiO in the low temperature regime can be written as
\begin{align} \label{eq:free_energy}
\mathcal{E}& = J \bm{m}_1 \cdot \bm{m}_2 - \frac{A}2 \left[(\bm{m}_1 \cdot \hat{x})^2 + (\bm{m}_2 \cdot \hat{x})^2\right] \notag\\
        &\quad + \frac{K}2 \left[(\bm{m}_1 \cdot \hat{z})^2 + (\bm{m}_2 \cdot \hat{z})^2\right] - M_s\bm{H} \cdot (\bm{m}_1+\bm{m}_2),
\end{align}
where $\bm{H}$ is the applied magnetic field; $J$, $A$ and $K$ are the exchange coupling, the easy-axis and the hard-axis anisotropy, respectively, all being positive in our convention. Without otherwise stated, thermal fluctuations will be ignored throughout this Letter. The coherent dynamics of $\bm{m}_1$ and $\bm{m}_2$ is described by the Laudau-Lifshitz-Gilbert (LLG) equations
\begin{subequations}\label{eq:LLG}
\begin{align}
&\dot{\bm{m}}_1 = \omega_J \bm{m}_{1} \times \bm{m}_{2} - \omega_A \bm{m}_{1} \times (\bm{m}_1 \cdot \hat{x})\hat{x} \notag\\ & + \omega_K \bm{m}_{1} \times (\bm{m}_1 \cdot \hat{z})\hat{z}   - \omega_H \bm{m}_{1} \times \hat{H} +  \alpha \bm{m}_{1} \times \dot{\bm{m}_{1}}, \\
&\dot{\bm{m}}_2 = \omega_J \bm{m}_{2} \times \bm{m}_{1} - \omega_A \bm{m}_{2} \times (\bm{m}_2 \cdot \hat{x})\hat{x} \notag\\ & + \omega_K \bm{m}_{2} \times (\bm{m}_2 \cdot \hat{z})\hat{z}   - \omega_H \bm{m}_{2} \times \hat{H} +  \alpha \bm{m}_{2} \times \dot{\bm{m}_{2}},
\end{align}
\end{subequations}
where $\hat{H}$ is the unit vector of $\bm{H}$, $\alpha$ is the Gilbert damping constant, $\omega_J=J/\hbar$, $\omega_A=A/\hbar$, $\omega_K=K/\hbar$ and $\omega_H=H M_s/\hbar$ are the angular frequencies corresponding to each magnetic interaction appearing in Eq.~\eqref{eq:free_energy}. When the system is driven by a microwave with an oscillating field $\bm{h}_{\rm{rf}}$, one should also add $-\bm{m}_{1,2}\times\bm{h}_{\rm{rf}}$ terms to the LLG equations, which will be discussed later.

\begin{figure}[t]
    \includegraphics[width = \linewidth]{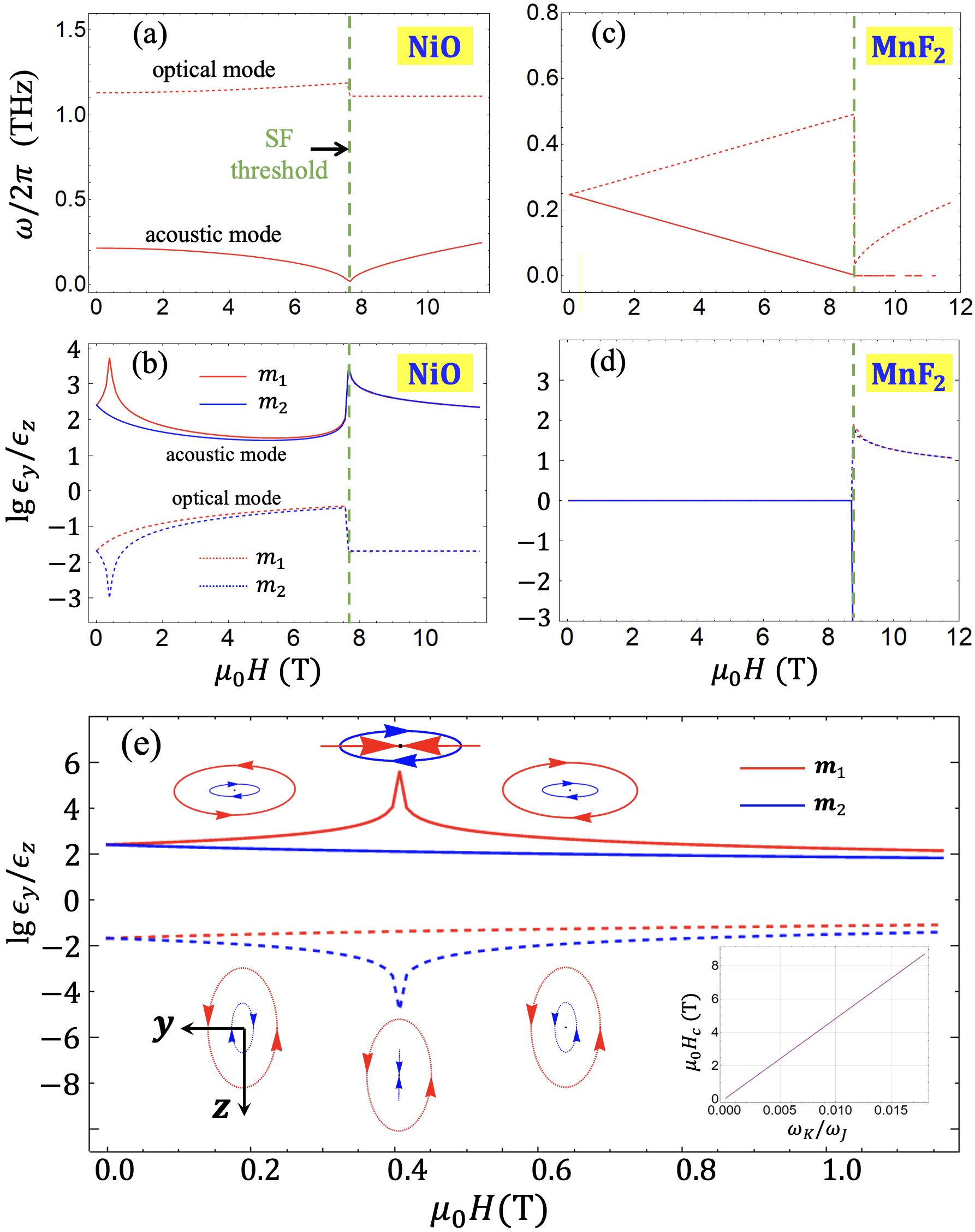}
    \caption{(a) eigenfrequencies and (b) polarization of the eigenmodes for NiO, compared with (c) and (d)---their counterparts in MnF$_2$. (e) Zoom-in plot of (b) in the vicinity of the critical point with illustrations of precession trajectories. Inset: dependence of the critical field on the hard-axis anisotropy. Parameters: for NiO, $\omega_J=1.7 \times 10^{13}$, $\omega_K=1.4\times 10^{10}$, $\omega_A=5.2\times 10^{8}$ (rad/s)~\cite{machado2017spin}, and $\alpha=5\times10^{-4}$~\cite{moriyama2019intrinsic}; for MnF$_2$, $\omega_K=0$ and other parameters taken from Ref.~\cite{vaidya2020subterahertz,nagamiya1955antiferromagnetism}}
    \label{fig:eigenmode}
\end{figure}

For $\bm{H}$ applied to the $\hat{x}$ direction (parallel to $\bm{m}_1$), we linearize the LLG equations around the equilibrium configuration, $\bm{m}_{1,2}=\pm\hat{x}$, and solve the eigenfrequencies adopting material parameters of NiO~\cite{machado2017spin,rezende2019introduction}. Figure~\ref{fig:eigenmode}(a) shows the eigenfrequencies of NiO, where the gap at zero field is attributed to the hard-axis anisotropy $\omega_K$. The two eigenfrequencies, dubbed acoustic and optical modes, vary non-linearly but continuously with an increasing field until the SF transition marked by the dashed green line around $7.8$T. In Fig.~\ref{fig:eigenmode}(b), we plot the polarization of each eigenvector in terms of the logarithm to the base $10$ of the ratio of principal axes of the elliptical trajectories for $\bm{m}_1$ (red lines) and $\bm{m}_2$ (blue lines), where we observe an unexpected divergence at around $0.4$T---far below the SF threshold. Specifically, in the acoustic (optical) mode with relatively low (high) frequency, it is $\bm{m}_1$ ($\bm{m}_2$) that experiences a diverging ratio $\epsilon_y/\epsilon_z$ at this critical point while the polarization of $\bm{m}_2$ ($\bm{m}_1$) changes smoothly.

As a comparison, we also show the eigenfrequencies and the polarization of the eigenmodes in uniaxial AFM MnF$_2$~\cite{vaidya2020subterahertz,nagamiya1955antiferromagnetism} in Fig.~\ref{fig:eigenmode}(c) and (d). On the one hand, the two eigenfrequencies are degenerate at zero field and splits linearly with an increasing field until the acoustic branch almost touches zero. On the other hand, both modes are circularly polarized below the SF transition as $\mathrm{lg} (\epsilon_y/\epsilon_z)=0$~\cite{keffer1952theory}.

To uncover the physical picture behind the unexpected divergence appearing in Fig.~\ref{fig:eigenmode}(b), we zoom in around the critical point and schematically illustrate the evolution of the precession trajectories of the two anti-parallel magnetic moments in Fig.~\ref{fig:eigenmode}(e). For the acoustic branch, $\bm{m}_1$ and $\bm{m}_2$ both rotate elliptically as seen against the $x$ axis, where the major (minor) axis is along $\hat{y}$ ($\hat{z}$). As the field goes across the critical point at which $\epsilon_y/\epsilon_z\rightarrow\infty$, the trajectory of $\bm{m}_1$ shrinks and reopens with an opposite chirality whereas $\bm{m}_2$ does not change its chirality. The optical branch follows a somewhat similar pattern but the polarization changes in the opposite way. Basing on Fig.~\ref{fig:eigenmode}(e), we identify a wide window of field strength between the critical point and the SF threshold within which the two magnetic moments precess with the same chirality---a crucial behavior for nonzero DC spin pumping~\cite{cheng2014spin}. Moreover, we find that the critical field $H_c$ for chirality flip increases linearly with the hard-axis anisotropy $\omega_K$ as shown in the inset of Fig.~\ref{fig:eigenmode}(e).

Specifically, if the two magnetic moments precess with opposite chirality, they will destructively contribute to the DC spin pumping as $\bm{m}_1\times\dot{\bm{m}}_1$ and $\bm{m}_2\times\dot{\bm{m}}_2$ are opposite in direction~\cite{cheng2014spin}. The two sublattice moments become constructive in generating spin current only when they rotate in the same manner. Therefore, the occurrence of chirality flip below the SF transition could enable the realization of DC spin pumping in NiO, hence in similar easy-plane AFM materials even within the collinear phase. This field-induced chirality flip, which has been overlooked in previous studies, reflects the restoration of rotational symmetry about the easy axis arising from the Zeeman interaction. For example, in the acoustic mode, the precession of $\bm{m}_1$ (red) gradually evolves from one chirality to the opposite chirality in order to minimize the energy penalty incurred from the increasing magnetic field. The critical point $H_c$ is where the chirality flips sign.

\begin{figure}[t]
    \includegraphics[width = 0.9\linewidth]{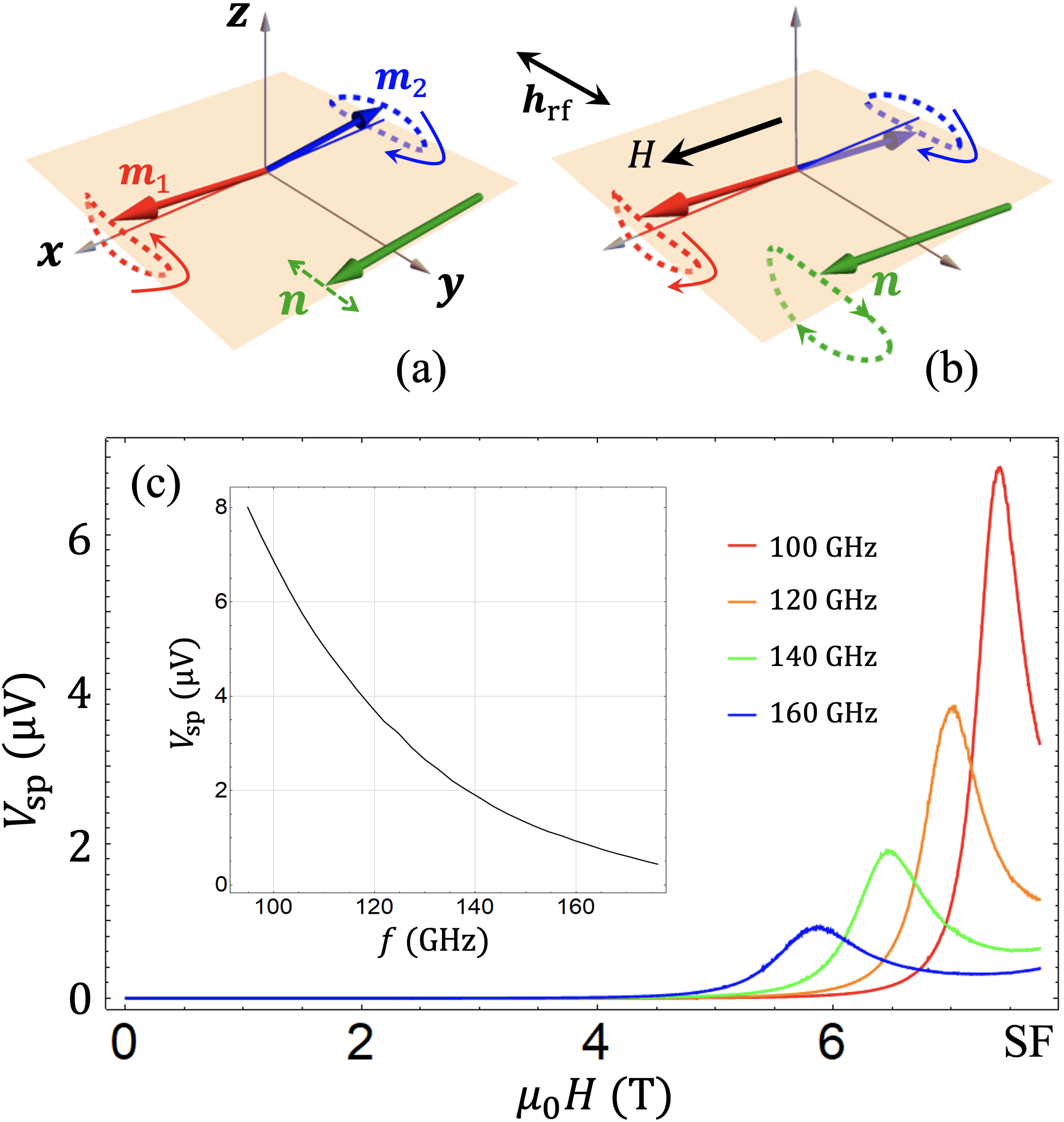}
    \caption{(a) and (b): Schematics of the acoustic mode in NiO for zero field and non-zero field surmounting the critical point of chirality flip. The trajectories are exaggerate for visual clarity. (c) Strength of DC spin pumping as a function of the applied magnetic field at different driving frequencies in the acoustic branch, following Eq.~\ref{equ:voltage_signal}. Inset: the output voltage varies with the driving frequency ranging from $90$ to $180$GHz. Parameters of the Pt detector assume values in Ref.~\cite{wang2021spin} while those for NiO taken from Ref.~\cite{machado2017spin}.}
    \label{fig:spin pumping}
\end{figure}

To demonstrate the field-assisted spin pumping more pictorially, we now investigate the dynamics of the acoustic mode of NiO in Fig.~\ref{fig:spin pumping}. In Fig~\ref{fig:spin pumping}(a), when there is no magnetic field, $\bm{m}_1$ (red) and $\bm{m}_2$ (blue) precess elliptically with opposite chirality, resulting in a linearly polarized Néel vector $\bm{n}=(\bm{m}_1-\bm{m}_2)/2$ (the trajectories have been exaggerated for visual clarity). Correspondingly, the resonance of $\bm{n}$ only give rise to an AC spin pumping while the DC component vanishes identically, which is a universal problem rooted in the broken rotational symmetry of bi-axial AFM materials. In addition, the total magnetic moment $\bm{m}=(\bm{m}_1+\bm{m}_2)/2$ oscillates linearly in the $z$ direction, so a driving microwave field $\bm{h}_{\rm{rf}}$ polarized along $y$ cannot even drive the acoustic mode. However, assisted by a Zeeman field, the trajectory of $\bm{m}_1$ flips its chirality across the aforementioned critical point, thereby leading to an elliptical precession of $\bm{n}$ as illustrated in Fig.~\ref{fig:spin pumping}(b). Moreover, once $\bm{n}$ opens up an elliptical trajectory, the total moment $\bm{m}$ will also rotate elliptically with a finite projection on the $y$ direction, which can couple to and be driven by the rf field. As such, a finite DC spin pumping is enabled, which can be detected straightforwardly by the inverse spin Hall effect (ISHE) in an adjacent heavy metal.

To quantify the physical picture described above, we now consider a Pt/NiO bilayer heterostructure and calculate the ISHE voltage arising from coherent DC spin pumping under a fixed driving frequency $\omega$ and a sweeping magnetic field $H$ along $\hat{x}$. For $\omega$ lying in the frequency range of the acoustic branch, the resonance field is at~\cite{machado2017spin,safin2022theory,safin2020electrically}
\begin{align}
  H_{\rm res} \approx \frac{\hbar}{M_s}\sqrt{2\omega_J\omega_A-\omega^2},
  \label{eq:resfreq}
\end{align}
which is true in the exchange limit that $\omega_A\ll\omega_K\ll\omega_J$. Basing on the geometry illustrated in Fig.~\ref{fig:spin pumping} and ignoring thermal excitations (\textit{e.g.}, spin Seebeck effect), we have the ISHE voltage, which is in the $y$ direction, as~\cite{wang2021spin}
\begin{equation}
 V_{\mathrm{sp}} = \omega \mathrm{Im}\left[\chi_{yy}^*\chi_{zy}\right] \tilde{g}_r \frac{eL\rho\theta_{SH}}{2\pi} \frac{\lambda}{d_N} \tanh\left(\frac{d_N}{2 \lambda}\right) |\bm{h}_{\rm{rf}}|^2,
 \label{equ:voltage_signal}
\end{equation}
where $\chi_{ij}=\chi_{ij}(\omega, H_{\mathrm{res}})$ ($i,j$ running over $y$ and $z$) is the dynamical susceptibility tensor defined as
\begin{align}
 n_i=\chi_{ij}(\omega, H_{\mathrm{res}})h_{\rm{rf},i},
\end{align}
which characterizes the response of the Néel vector to the microwave drive $\bm{h}_{\rm{rf}}$. In Eq.~\eqref{equ:voltage_signal}, $e$ is the electron charge, $L$, $\rho$, $\theta_{SH}$, $\lambda$ and $d_N$ are the length, resistivity, the spin Hall angle, spin diffusion length, and thickness of the Pt layer, respectively. $\tilde{g}_r$ is the real part of the spin-mixing conductance determining the efficiency of spin transmission across the Pt/NiO interface, which is typically affected by the spin backflow effect and the spin diffusion effect in the Pt~\cite{brataas2002spin,jiao2013spin}.

Figure~\ref{fig:spin pumping}(c) plots the numerical results based on Eq.~\ref{equ:voltage_signal} for four different driving frequencies in the acoustic branch, where a lower driving frequency results in a higher spin pumping signal taking place at a higher resonance field, consistent with Eq.~\eqref{eq:resfreq}. As $H_{\rm{res}}$ approaches the SF threshold, the polarization of sublattice magnetic moments, hence that of $\bm{n}$, becomes increasingly circular, greatly enhancing the strength of DC spin pumping. At about $100$ GHz, the amplitude of spin pumping is as large as that in a uniaxial AFM material. In the inset, we show the dependence of the ISHE voltage on the driving frequency ranging from $90$ GHz to $180$ GHz, which is an accessible range for microwave drive.

\begin{figure*}[t]
    \includegraphics[width = 0.8\linewidth]{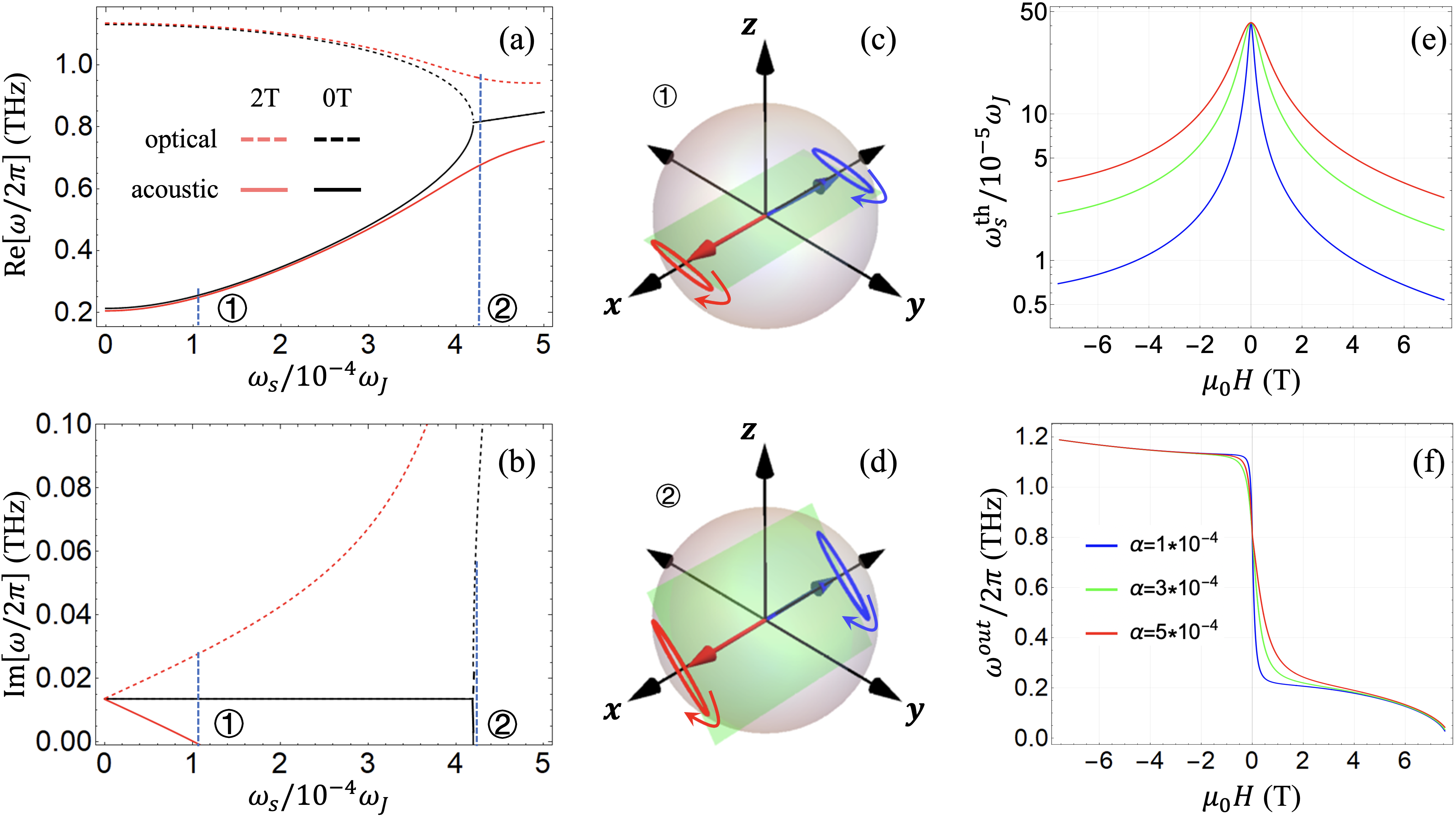}
    \caption{(a) Real and (b) imaginary parts of the eigenfrequencies of NiO as functions of the STT strength $\omega_s$ for zero field (black curves) and $\mu_0H=2$T (red curves). The threshold STT in the presence (absence) of magnetic field is marked by \ding{172} (\ding{173}). Note that \ding{173} is only slightly above than the bifurcation point. (c) and (d): schematics of the eigenmodes at thresholds \ding{172} and \ding{173}, respectively. (e) Threshold values of STT and (f) output frequency of the auto-oscillation as functions of the applied magnetic field for different Gilbert damping constants.}
    \label{fig:STT}
\end{figure*}

Given that a magnetic field can substantially compensate the broken symmetry ascribing to the bi-axial anisotropy to enable an appreciable DC spin pumping even in the collinear phase, it is natural to expect a similar remarkable field-assisted effect in the reciprocal phenomenon, namely the reduction of the current threshold triggering auto-oscillation of the N\'{e}el vector. To this end, we add the spin-transfer torque (STT)~\cite{slonczewski1996current,berger1996emission} $\bm{\tau}_i = \bm{m}_i \times (\bm{\omega}_s \times \bm{m}_i)$ to Eq.~\ref{eq:LLG}. Here, $\bm{\omega}_s=\omega_s\hat{x}$, and $\omega_s$ is the strength of STT scaled into the frequency dimension~\cite{cheng2015ultrafast,cheng2016terahertz}
\begin{align}
 \omega_s=J_c\frac{\theta_{SH}e\rho\tilde{g}_r\lambda a^3}{h d_{\rm NiO}},
\end{align}
where $J_c$ is the applied current density in the $y$ direction, $a$ and $d_{\rm NiO}$ are the lattice constant and thickness of the NiO layer. For $\bm{n} \approx \hat{x} + (n_y \hat{y} + n_z \hat{z}) \mathrm{e}^{\mathrm{i} \omega t} $ and $\bm{m} \approx (m_y \hat{y} + m_z \hat{z}) \mathrm{e}^{\mathrm{i} \omega t} $, where $n_{y,z}\ll1$ and $m_{y,z}\ll1$, linearizing the LLG equations with respect to the basis $\{m_y,m_z,n_y,n_z\}^{\mathrm{T}}$ results in an eigenvalue problem. The eigenfrequencies $\omega$ are determined by
\begin{align}
 &\begin{vmatrix}
  \mathrm{i}\omega & \omega_H & \omega_s & \omega_K+\omega_A+\mathrm{i}\alpha\omega \\
  -\omega_H & \mathrm{i}\omega & \omega_A+\mathrm{i}\alpha\omega & \omega_s \\
  0 & 2\omega_J+\omega_K+\omega_A & \mathrm{i}\omega & \omega_H \\
  -2\omega_J-\omega_A & 0 & -\omega_H & \mathrm{i}\omega
 \end{vmatrix} \notag\\
 &\qquad\qquad\qquad\qquad\qquad\qquad\qquad\qquad =0, \label{eq:eigenfreq}
\end{align}
where only two of the four solutions are physically distinct while the other two are redundant~\cite{keffer1952theory}. We label the two physical solutions as the acoustic mode $\omega_{ac}$ and the optical mode $\omega_{op}$. In the presence of the STT competing with the Gilbert damping, both $\omega_{ac}$ and $\omega_{op}$ are complex valued, where the real parts determine the precessional frequency and the imaginary parts determine the stability of the eigenmodes. If the imaginary part $\rm{Im}[\omega]$ turns zero, the corresponding eigenmode becomes unstable, marking the onset of auto-oscillation. Figure~\ref{fig:STT}(a) and (b) plot the real and imaginary parts for both modes as functions of the STT strength $\omega_s$ for $\mu_0H=0$ T (black curves) and $\mu_0H=2$T (red curves) applied in the same direction as $\bm{\omega}_s$. Note that a $2$T magnetic field already surpasses the critical field so that the N\'{e}el vector acquires the left-handed chirality in $\omega_{ac}$.

For $\mu_0H=0$T, $\mathrm{Re}[\omega_{ac}]$ and $\mathrm{Re}[\omega_{op}]$ approach each other as the STT increases until they merge at a bifurcation point, while $\mathrm{Im}[\omega_{ac}]$ and $\mathrm{Im}[\omega_{op}]$ are degenerate and kept constant. Beyond the bifurcation point, $\mathrm{Im}[\omega_{ac}]$ rapidly decreases and touches zero at \ding{173}, which is only slightly above the bifurcation point~\cite{notealpha}. According to Ref.~\cite{cheng2016terahertz}, the anti-damping effect of STT only takes place within the narrow window bounded by the bifurcation point and \ding{173}, while the vast majority of the STT counteracts the impact of the hard-axis anisotropy rather than competes directly with the Gilbert damping.

For $\mu_0H=2$T, the situation changes dramatically. The bifurcation point of the imaginary parts is moved to zero, so applying an STT immediately exerts an anti-damping effect on the N\'{e}el vector, and the auto-oscillation of the acoustic mode is triggered at a much lower threshold marked as \ding{172}. For the real parts, the bifurcation point is also removed (in fact, be moved to infinity) by the Zeeman field such that the two modes never merge. Therefore, the field-assisted effect manifests as the elimination of the bifurcation point, hence a significant reduction of the auto-oscillation threshold where $\rm{Im}[\omega_{ac}]=0$.

Having obtained the influence of Zeeman field on the eigenfrequencies, it is natural to check the eigenvectors from which we can draw a more intuitive picture of the auto-oscillation. What is relevant to our discussion is the way magnetic moments precess around their equilibrium directions right at the auto-oscillation threshold (\textit{i.e.}, $\rm{Im}[\omega_{ac}]=0$). To this end, we numerically simulate the eigenvectors at \ding{172} (threshold for $\mu_0 H=2$T) and \ding{173} (threshold for zero field). Figure~\ref{fig:STT}(c) and~(d) show their snapshots with amplitude exaggerated for visual clarity. While they both exhibit the same chirality (left-handed) for both sublattice magnetic moments, the polarization plane enclosing the major axis of elliptical trajectories (highlighted in green) is different. The tilting angle of the polarization plane with respect to the easy plane of NiO is negligibly small for \ding{172} and about $45^\circ$ for \ding{173}. In the latter, the apparent inclination is a manifestation of the counteraction of hard-axis anisotropy by the STT~\cite{cheng2016terahertz}.

We emphasize that the Zeeman field does not directly lead to instability. It is the anti-damping effect of the STT that triggers the auto-oscillation. However, the anti-damping effect is inhibited remarkably by the hard-axis anisotropy of NiO in the absence of magnetic fields. What a Zeeman field can do is to restore the broken rotational symmetry by suppressing the hard-axis anisotropy $\omega_K$, thereby assisting the STT to overcome the Gilbert damping (\textit{i.e.}, anti-damping effect).

Finally, we quantify the field-assisted auto-oscillation by calculating the threshold STT $\omega_s^{\rm{th}}$ as a function of the applied magnetic field for different Gilbert damping constants. As shown in Fig.~\ref{fig:STT}(e), the threshold STT $\omega_s^{\rm{th}}$ decreases monotonically with an increasing magnetic field; this field-assisted reduction is more compelling for smaller Gilbert damping. For example, using the experimental value $\alpha= 5 \times 10^{-4}$~\cite{moriyama2019intrinsic}, we estimate that a magnetic field of $6$T can reduce $\omega_s^{\rm{th}}$ by more than one order of magnitude (red curve). For even smaller damping, \textit{e.g.} $\alpha=1\times10^{-4}$, $\omega_s^{\rm{th}}$ is reduced by almost two orders of magnitude at $6$T. Moreover, $\omega_s^{\rm{th}}$ is an asymmetric function of $H$: the reduction is more prominent for positive $H$ (when Zeeman field is parallel to the current-induced spin accumulation $\bm{\omega}_s$). This asymmetry is reflected more clearly in the output auto-oscillation frequency as plotted in Fig.~\ref{fig:STT}(f). Specifically, when the magnetic field is parallel (antiparallel) to the spin accumulation $\bm{\omega}_{s}$, the acoustic (optical) mode will be excited by the STT. However, when the optical mode is driven into auto-oscillation (for negative $H$), the system cannot stabilize on itself and the precessional motion will inevitably evolve into a large-angle precession in the $y-z$ plane. Unlike the threshold value of STT, the output frequency as a function of $H$ is almost independent of $\alpha$ above $2$T ($1$T) for positive (negative) magnetic field, as shown Fig.~\ref{fig:STT}(f). This is because once an auto-oscillation is triggered, its output frequency is determined by the real part of the eigenfrequency that is insensitive to the Gilbert damping.

In both the spin pumping and the spin-torque oscillator studied above, if the magnetic field is applied to another direction, the overall effects become smaller because: 1) it is the Zeeman field projected onto the Néel vector that changes the eigenmode polarization; 2) if the Néel vector deviates from the easy axis while $\bm{h}_{\rm rf}$ does not change then the microwave absorption becomes weaker.

\begin{acknowledgments}
This work is supported by the Air Force Office of Scientific Research under grant FA9550-19-1-0307.
\end{acknowledgments}


\begin{thebibliography}{26}

    \bibitem{baltz2018antiferromagnetic} V. Baltz, A. Manchon, M. Tsoi, T. Moriyama, T. Ono, and Y. Tserkovnyak, “Antiferromagnetic spintronics,” Reviews of Modern Physics 90, 015005 (2018)
    
	\bibitem{jungwirth2018multiple} T. Jungwirth, J. Sinova, A. Manchon, X. Marti, J. Wunderlich, and C. Felser, “The multiple directions of antiferromagnetic spintronics,” Nature Physics 14, 200–203 (2018).
	
	\bibitem{gomonay2014spintronics} E. Gomonay and V. Loktev, “Spintronics of antiferromagnetic systems,” Low Temperature Physics 40, 17–35 (2014).
	
	\bibitem{cheng2014spin} R. Cheng, J. Xiao, Q. Niu, and A. Brataas, “Spin pumping and spin-transfer torques in antiferromagnets,” Physical review letters 113, 057601 (2014).
	
	\bibitem{johansen2017spin} Ø. Johansen and A. Brataas, “Spin pumping and inverse spin Hall voltages from dynamical antiferromagnets,” Physical Review B 95, 220408 (2017).
	
	\bibitem{vaidya2020subterahertz} P. Vaidya, S. A. Morley, J. van Tol, Y. Liu, R. Cheng, A. Brataas, D. Lederman, and E. Del Barco, “Subterahertz spin pumping from an insulating antiferromagnet,” Science 368, 160–165 (2020).
	
	\bibitem{li2020spin} J. Li, C. B. Wilson, R. Cheng, M. Lohmann, M. Kavand, W. Yuan, M. Aldosary, N. Agladze, P. Wei, M. S. Sherwin, et al., “Spin current from sub-terahertz-generated antiferromagnetic magnons,” Nature 578, 70–74 (2020).
	
	\bibitem{sievers1963far} A. Sievers III and M. Tinkham, “Far infrared antiferromagnetic resonance in MnO and NiO,” Physical Review 129, 1566 (1963).
	
	\bibitem{wang2021spin} H. Wang, Y. Xiao, M. Guo, E. Lee-Wong, G. Q. Yan, R. Cheng, and C. R. Du, “Spin Pumping of an Easy-Plane Antiferromagnet Enhanced by Dzyaloshinskii–Moriya Interaction,” Physical Review Letters 127, 117202 (2021).
	
	\bibitem{boventer2021room} I. Boventer, H. T. Simensen, A. Anane, M. Kläui, A. Brataas, and R. Lebrun, “Room-temperature antiferromagnetic resonance and inverse spinHall voltage in canted antiferromagnets,” Physical review letters 126, 187201 (2021).
	
	\bibitem{williamson1964AFM} 1S. J. Williamson and S. Foner, “Antiferromagnetic Resonance in Systems with Dzyaloshinsky-Moriya Coupling; Orientation Dependence in $\alpha-$Fe$_2$O$_3$,” Physical Review 136, A1102 (1964).
	
	\bibitem{cheng2016terahertz} R. Cheng, D. Xiao, and A. Brataas, “Terahertz antiferromagnetic spin hall nano-oscillator,” Physical review letters 116, 207603 (2016)

	\bibitem{khymyn2017AFM} R. Khymyn, I. Lisenkov, V. Tiberkevich, B. A. Ivanov, and A. Slavin, “Antiferromagnetic THz-frequency Josephson-like oscillator driven by spin current,” Scientific reports 7, 1–10 (2017).
	
	\bibitem{machado2017spin} F. Machado, P. Ribeiro, J. Holanda, R. Rodríguez-Suárez, A. Azevedo, and S. Rezende, “Spin-flop transition in the easy-plane antiferromagnet nickel oxide,” Physical Review B 95, 104418 (2017).
	
	\bibitem{rezende2019introduction} S. M. Rezende, A. Azevedo, and R. L. Rodríguez-Suárez, “Introduction to antiferromagnetic magnons,” Journal of Applied Physics 126, 151101 (2019).

	\bibitem{nagamiya1955antiferromagnetism} T. Nagamiya, K. Yosida, and R. Kubo, “Antiferromagnetism,” Advances in Physics 4, 1–112 (1955).
	
	\bibitem{keffer1952theory} F. Keffer and C. Kittel, “Theory of antiferromagnetic resonance,” Physical Review 85, 329 (1952).
	
	\bibitem{moriyama2019intrinsic} T. Moriyama, K. Hayashi, K. Yamada, M. Shima, Y. Ohya, and T. Ono, “Intrinsic and extrinsic antiferromagnetic damping in NiO,” Physical Review Materials 3, 051402 (2019).
	
	\bibitem{safin2022theory} A. Safin, S. Nikitov, A. Kirilyuk, V. Tyberkevych, and A. Slavin, “Theory of antiferromagnet-based detector of terahertz frequency signals,” Magnetochemistry 8, 26 (2022).
	
	\bibitem{safin2020electrically} A. Safin, V. Puliafito, M. Carpentieri, G. Finocchio, S. Nikitov, P. Stremoukhov, A. Kirilyuk, V. Tyberkevych, and A. Slavin, “Electrically tunable detector of THz-frequency signals based on an antiferromagnet,” Applied Physics Letters 117, 222411 (2020).

	\bibitem{brataas2002spin} A. Brataas, Y. Tserkovnyak, G. E. Bauer, and B. I. Halperin, “Spin battery operated by ferromagnetic resonance,” Physical Review B 66, 060404 (2002).
	
	\bibitem{jiao2013spin} H. Jiao and G. E. Bauer, “Spin backflow and ac voltage generation by spin pumping and the inverse spin Hall effect,” Physical review letters 110, 217602 (2013).
	
	\bibitem{slonczewski1996current} J. C. Slonczewski, “Current-driven excitation of magnetic multilayers,” Journal of Magnetism and Magnetic Materials 159, L1–L7 (1996).
	
	\bibitem{berger1996emission} L. Berger, “Emission of spin waves by a magnetic multilayer traversed by a current,” Physical Review B 54, 9353 (1996).
	
	\bibitem{cheng2015ultrafast} R. Cheng, M. W. Daniels, J.-G. Zhu, and D. Xiao, “Ultrafast switching of antiferromagnets via spin-transfer torque,” Physical Review B 91, 064423 (2015).
	
	\bibitem{notealpha} Note that in Ref.~\cite{cheng2016terahertz} we overestimated $\alpha$ so that the auto-oscillation threshold \ding{173} and the bifurcation point are not too close to each other. However, a more recent experiment~\cite{moriyama2019intrinsic} shows that $\alpha$ is only about $5\times10^{-4}$, which we adopted in this Letter, thus \ding{173} is only slightly above the bifurcation point.

\end{thebibliography}
\end{document}